\documentclass[fontsize = 11pt ,
paper =a4, 
draft =false , 
titlepage=false , 
twoside =false , 
parskip=half ]{scrartcl}
\usepackage[utf8]{inputenc}
\usepackage[T1]{fontenc}
\usepackage[english]{babel}
\usepackage{amsmath}
\usepackage{dsfont}
\usepackage{amsfonts}
\usepackage{bm}
\usepackage{csquotes}
\usepackage{tikz}
\usepackage{subcaption}
\usepackage{booktabs}
\usepackage{array}
\usepackage{enumitem}
\usepackage{adjustbox}
\usepackage{multirow}
\usepackage{tabularx}
\usepackage{authblk}
\usepackage{hyperref}
\usepackage{orcidlink}
\usepackage[defaultlines = 3, all]{nowidow}
\usepackage[dvipsnames]{xcolor}

\usepackage{setspace}
\usepackage{anysize}
\marginsize{1.9cm}{1.9cm}{1.9cm}{1.9cm}

\usepackage{hyperref}
\usepackage{xurl}
\usepackage{graphicx}
\usepackage{amssymb}

\usepackage[backend=biber, 
			style=apa, 
			citestyle = authoryear-comp, 
			maxcitenames=2, 
			maxbibnames=20, 
			dashed = false, 
			giveninits=true, 
			url = true, 
			natbib=true, 
			doi = true, 
			eprint =false, 
			bibstyle =authoryear, 
			isbn=false, clearlang=true, 
            uniquename=false,
			uniquelist=false]{biblatex}

\renewbibmacro*{volume+number+eid}{
	\printfield{volume}
	\setunit*{\addnbthinspace}
	\printfield{number}
    \setunit{\addcomma\space}
	\printfield{eid}}
\DeclareFieldFormat[article]{number}{\mkbibparens{#1}}
\DeclareFieldFormat[article]{eid}{#1}

\AtEveryBibitem{\clearfield{month}}
\AtEveryCitekey{\clearfield{month}}
\AtEveryBibitem{\clearfield{day}}
\AtEveryCitekey{\clearfield{day}}
\AtEveryBibitem{\clearfield{note}}
\AtEveryCitekey{\clearfield{note}}

\DefineBibliographyStrings{german}{andothers={\mbox{et~al.}}}

\providecommand{\keywords}[1]
{
  \small	
  \textbf{\textit{Keywords---}} #1
}

\NewBibliographyString{toappear}
\DefineBibliographyStrings{english}{%
  toappear = {to appear in:},
  version = {R package version},
}

\renewbibmacro*{in:}{%
  \iffieldundef{pubstate}
    {}
    {\printfield{pubstate}%
     \setunit{\addspace}%
     \clearfield{pubstate}}%
    }

\usepackage{breakurl}

\newcolumntype{C}[1]{>{\centering\let\newline\\\arraybackslash\hspace{0pt}}m{#1}}

\usepackage{lmodern} 
\usepackage{relsize}

\setkomafont{section}{\rmfamily\bfseries\normalsize}
\setkomafont{subsection}{\rmfamily\bfseries\normalsize}
\setkomafont{subsubsection}{\rmfamily\bfseries\normalsize}
\begin{document}
    \title{ \vspace{-1em}\Large{\textrm{Modeling inter-exceedance times of extreme heat days}}}
    \author[1, 3]{Merle Mendel \orcidlink{0009-0006-7817-2963}}
    \author[1]{Roland Fried \orcidlink{0000-0002-9830-9713}}
    \author[1, 2]{Nils Weitzel \orcidlink{0000-0002-2735-2992}}
    \affil[1]{Department of Statistics, TU Dortmund University, Germany}
    \affil[2]{Research Center Trustworthy Data Science and Security of the University Alliance Ruhr, Dortmund, Germany}
    \affil[3]{Corresponding author: merle.mendel@tu-dortmund.de}
	\maketitle

    \begin{center}
    \textbf{Abstract}
\end{center}

\small Extreme heat days, here defined as exceedances of a local calendar-day-based 95\% quantile, often occur in clusters forming multi-day heatwaves. A calendar-day based threshold allows for the detection of temperature extremes across all seasons, which may impact several economic sectors, public health and ecosystems. Previous approaches to modeling the occurrences of extreme heat events typically treat heat event duration and the time between heat events separately. We propose modeling the inter-exceedance times (IETs) between extreme heat days, allowing for a unified description within a single model. We consider several candidate distributions for the IETs of extreme heat days and find that the zero-inflated discrete Weibull (ZIDW) distribution clearly provides the best fit according to Pearson's chi-squared goodness-of-fit tests. The ZIDW distribution is a flexible model for count data exhibiting excess zeros as well as over- and under-dispersion: the zero-inflation component captures the pronounced clustering of heat days, while the discrete Weibull component adequately accounts for over-dispersion. Heat days are extracted using both trend-adjusted and non–trend-adjusted thresholds. This allows the analysis of general changes in the occurrence of extreme heat days as well as changes after accounting for the general increase in temperature. 


	\section{Introduction}
Extreme heat events have severe impacts on the environment, agriculture, and human health. Recent examples include the 2024 European heatwaves, during which more than 62,000 deaths related to heat occurred as estimated by \citet{example_euro24}.
For 2026, the Robert Koch Institute estimates that around 16,000 heat-related deaths occurred in Germany alone \citep{rki_report}, exceeding estimates from previous years.
Outside Europe, Southeastern Australia experienced a severe heatwave in January 2026, during which bushfires destroyed hundreds of homes and other buildings \citep{bbc_example, example_australia2}. 

There is no single definition of extreme heat events or heatwaves; instead, the definition is often tailored to the type of damage of interest. If the focus primarily lies on human health, a fixed temperature threshold for the daily temperature maxima is commonly used, as in \citet{Abaurrea_2007} or \citet{Reich_2014}, who consider the local 95\% quantile.
Local quantile thresholds are often preferred to absolute thresholds, since regions can adapt to the local climate \citep{Winter_2017}. Frequently, a temperature threshold is additionally paired with a minimum duration. For instance, \citet{stefano_2012} require a minimum of four consecutive days above the locally defined 95\% quantile. 
If the focus lies on the ecosystem instead, a time-varying threshold is more appropriate, as temperature extremes away from summer affect agriculture and environment. To implement this, \citet{Russo_2014} use quantile thresholds based on calendar-days. In this study, we adopt this approach to enable year-round analyses of temperature extremes. Due to climate warming, heat extremes are becoming more frequent and intense \citep{emerton_26}.
To account for this, we consider both trend-adjusted and non–trend-adjusted thresholds. The trend-adjusted threshold allows changes in the behavior of heat events to be analyzed independently of the general increase in temperature, whereas a non–trend-adjusted threshold permits investigating overall changes in heat event behavior (e.g., \cite{dillerup_2026}).
\citet{Abaurrea_2007} define consecutive extreme heat days as a heatwave, and they model the heatwave occurrences with a non-homogeneous Poisson process. The heatwave duration is then modeled separately by a shifted negative binomial distribution. In contrast, we propose modeling the inter-exceedance times (IETs) between individual extreme heat days. Hence, the duration does not need to be modeled separately in this approach, as the model allows for the direct succession of extreme heat days.
The IETs show over-dispersion, meaning that the variance is larger than the mean, which also needs to be taken into account when selecting a statistical model. Additionally, mid-latitude heat waves are typically associated with persistent atmospheric blocking \citep{Kautz_2022}, resulting in frequent occurrences of consecutive heat days and hence requiring a model capable of capturing temporal clustering. 

In this study, we consider several candidate distributions for the IETs of extreme heat days, including the zero-inflated negative binomial or zero-inflated geometric distribution. The zero-inflated discrete Weibull (ZIDW) distribution clearly provides the best fit according to Pearson's chi-squared goodness-of-fit tests. Moreover, modeling the IETs between extreme heat days using the ZIDW distribution also provides a good representation of heatwave durations. We use the ZIDW distribution to analyze temporal changes in the behavior of extreme heat days stratified by season. This is done for data derived from both non–trend-adjusted and trend-adjusted thresholds, allowing for complementary perspectives. Furthermore, we evaluate the severity of the European summer 2026 based on heatwave durations.

\section{Methodology} \label{sec:methods1}
In this section, we review the discrete Weibull and zero-inflated discrete Weibull distributions and outline the corresponding regression framework.
The R-package \texttt{ZIDW} \citep{ZIDW_R} is used to fit the ZIDW distribution and the ZIDW regression models with maximum likelihood estimation numerically using the Nelder-Mead method.

\subsection{The discrete Weibull and zero-inflated discrete Weibull distributions}
The discrete Weibull distribution was first introduced by \citet{discrete_weibull} in the context of failure time analysis.
The cumulative distribution function (cdf) is given by
\[F(y) = \begin{cases}
    1-q^{(y+1)^\beta} , \quad \text{for }y\in\mathbb{N}_0;  \\
    0,  \quad \quad \quad\quad \quad \text{otherwise,}
\end{cases} \]
with shape parameters $q \in (0,1)$ and $\beta>0$. This results in the probability mass function (pmf)
\[f(y) = \begin{cases}
    q^{y^\beta}- q^{(y+1)^\beta}, \quad \text{for }y\in\mathbb{N}_0;  \\
    0,  \quad \quad \quad\quad \quad \;\;\;\text{otherwise.}
\end{cases} \]
For $\beta = 1$, the discrete Weibull distribution reduces to the geometric distribution with success probability $1-q$. An advantage of the discrete Weibull distribution is that it can handle over- and underdispersion relative to the Poisson, making it quite flexible.

A popular approach to address excess zeros in count data is zero inflation, where a point mass at zero is mixed with a discrete distribution.
This has been done for many different discrete distributions, starting with the zero-inflated Poisson distribution introduced in \citet{lambert_1992}. 
Recently, the zero-inflated discrete Weibull distribution (ZIDW) has been discussed in-depth by \citet{ZIDW}. The distribution accounts for excess zeros through a zero-inflation parameter $\lambda \in [0, 1]$ and has the pmf
\[f(y) =
    \lambda \mathds{1}_0(y)+ (1-\lambda)\left(q^{y^\beta}- q^{(y+1)^\beta}\right), \quad \text{for }y\in\mathbb{N}_0. \]
For the ZIDW, the probability of zero is equal to $f(0) = \lambda + (1-\lambda)(1-q) = 1-q +\lambda q$, which is larger by $\lambda q $ than in the discrete Weibull distribution.

\subsection{ZIDW regression}

ZIDW regression has recently been discussed by \citet{yeh_2023}. It is assumed that the response variable $Y$ follows a ZIDW distribution with density \[f\left(y_i; \lambda(x_i), q(w_i), \beta(z_i)\right),\]
where the parameters of the ZIDW distribution, $\lambda(x_i)$, $q(w_i)$, and $\beta(z_i)$, depend on explanatory variables through the link functions
\[\text{logit}(\lambda) = \log\left( \frac{\lambda}{1-\lambda}\right) = X \alpha, \quad \text{logit}(q) = \log\left( \frac{q}{1-q}\right) = W \zeta \quad \text{and} \quad \log(\beta) =  Z \gamma.\]
$X, W$ and $Z$ denote matrices of the explanatory variables and $\alpha, \zeta $ and $\gamma$ denote parameter vectors.
Hence, the ZIDW parameters $\lambda$, $q$, and $\beta$ can depend on different sets of explanatory variables.

    \section{Application to extreme heat days}\label{sec:example}

Given that extreme heat is typically associated with persistent weather regimes such as atmospheric blocking \citep{Kautz_2022}, extreme heat days tend to cluster into multi-day heatwaves.
In statistical modeling, heatwaves are often described using two separate models, one for the heatwave duration and another for the inter-exceedance times (IETs) between heatwaves; see, e.g., \citet{Abaurrea_2007}. 
In this section, we discuss an alternative approach by modeling the IETs between heat days. This framework allows us to jointly describe heatwave duration and the IETs between heatwaves within a single model. In Section~\ref{goodness-of-fit}, we assess the goodness-of-fit of several candidate distributions to the IETs, where the ZIDW distribution clearly provides the best results. In Section~\ref{analysis_trend}, we then use the ZIDW distribution to investigate temporal changes in different characteristics of the IETs. 

\subsection{Dataset}
\begin{figure}[!htbp]
    \centering
    \includegraphics[width = 0.65\linewidth]{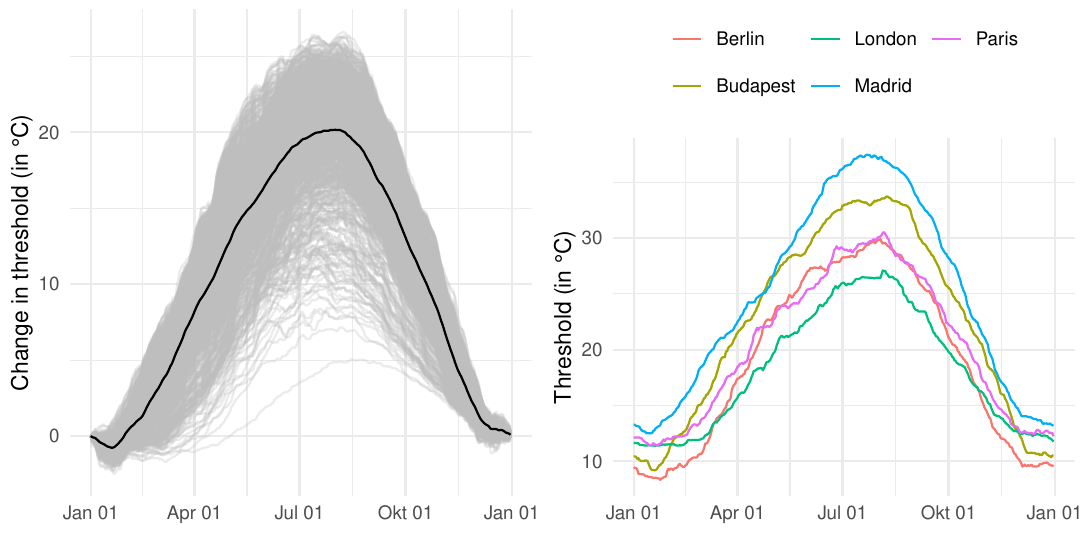}
    \caption{Left: Differences between the threshold and the threshold on January~1; individual grid points are shown in gray, the overall mean difference across the considered area is displayed in black. Right: Example thresholds for selected cities.}
    \label{threshold_seasonal}
\end{figure}

In the following, we analyze the daily maxima of the \mbox{2\,m} temperature of the ERA5 reanalysis data \citep{datensatz2}, provided by the European Centre for Medium-Range Weather Forecasts (ECMWF). 
We examine large parts of northern Africa and Europe (\mbox{30°~N\,–\,60°~N,} \mbox{10°~W\,–\,30°~E}) on a \mbox{1°\,$\times$\,1°} grid from January~1, 1960 to December~31, 2024. Additionally, we consider the same region from June~1 to August~31, 2026, to conduct an out-of-sample evaluation of summer 2026 in Section~\ref{2026}. Grid points are included in the analysis if their coordinates intersect land polygons from the Natural Earth dataset at medium scale \citep{naturalearth_countries}, resulting in a total of 758 grid points.
We apply the peaks over threshold method to extract heat days from the daily temperature maxima. To address the strong seasonality present in temperature data, we consider the calendar-day-based moving 95\%~quantile, centered on a 15-day window, as time-dependent threshold for each grid point separately.
A similar approach has been applied in \citet{Russo_2014}, who use a calendar-day-based 0.9 quantile of the daily temperature maxima and a 31-day window.
The resulting thresholds are illustrated in Figure~\ref{threshold_seasonal}. For comparison, the threshold on January~1 is subtracted from the thresholds at each location (left panel) and the mean threshold over all locations is shown in black. The right panel shows non-shifted example thresholds for selected European cities.

\begin{figure}[!htbp]
    \centering
    \includegraphics[width = 0.55\linewidth]{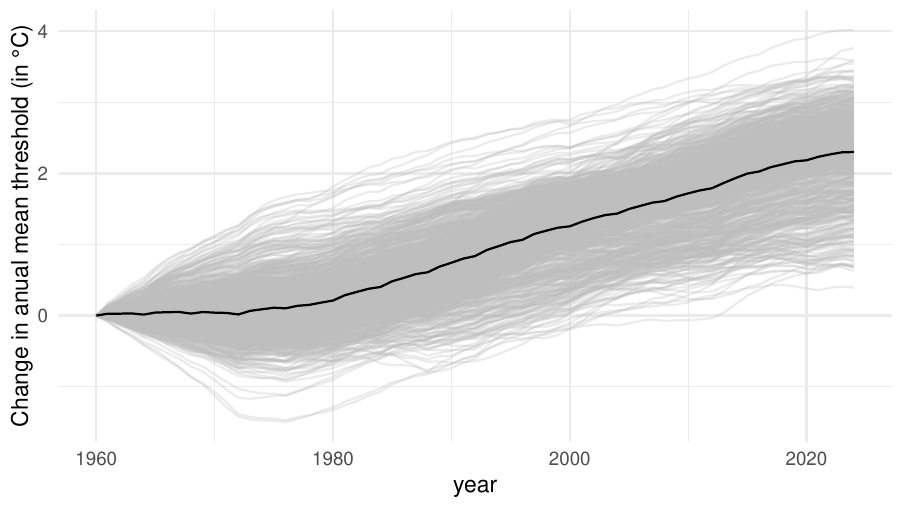}
    \caption{Difference between the annual mean trend-adjusted threshold and the mean trend-adjusted threshold of 1960; individual grid points are shown in gray, the overall mean difference across the considered area is displayed in black.}
    \label{threshold_epanechnikov}
\end{figure}

In addition, we consider thresholds that incorporate long-term trends at each grid point. To this end, we calculate the daily weighted moving 95\%~quantile using only observations within a 21-year window and a 15-day window centered around the respective calendar day. Observation weights are computed using an Epanechnikov kernel \citep{epanechnikov}, so that years farther from the center of the window receive less weight. The weighted quantile is computed as described in \citet{merle}. It should be noted that for the years 1970-2014 a full 21-year window centered on each year is available, whereas for the years 1960-1969 and 2015-2024 the window is truncated on one side due to limited data availability.   
The resulting trends in the thresholds are visualized in Figure~\ref{threshold_epanechnikov}, which depicts the difference between the annual mean threshold and the mean threshold in 1960 for all 758 grid points (gray). Additionally, the average across all grid points is shown in black. The trend-adjusted threshold allows assessing if and how the clustering behavior of extreme heat days changes when adjusting for the general increase in temperature. The ERA5 reanalysis data used in this study are available via the link provided in the Data Availability Statement.

\subsection{Goodness of fit evaluation}\label{goodness-of-fit}

To assess the goodness of fit of candidate distributions to the IETs between extreme heat days, we conduct Pearson's chi-squared goodness-of-fit tests. Specifically, we test
\[\boldsymbol{H_0}: \text{the IETs follow the candidate distribution} \]
at level 0.05. For these tests, the support of the IETs (i.e., $\mathbb{N}_0$) is partitioned into bins constructed so that the expected number of observations in each bin is at least five.
In addition to the ZIDW distribution, we consider four alternative zero-inflated distributions: the zero-inflated negative binomial (ZINB), zero-inflated generalized Poisson (ZIGP), zero-inflated geometric (ZIG) and zero-inflated Poisson (ZIP) distributions. The ZINB and ZIGP offer increased flexibility, relative to the ZIG and ZIP distributions, respectively, due to their additional parameters.  
We also include the discrete Weibull (DW) distribution, which does not explicitly account for excess zeros, as well as the continuous analog of the ZIDW distribution, defined as a mixture of a Weibull distribution and a point mass at zero ($\delta_0$+WB).
Finally, since the fractional compound Poisson process (FCPP) has previously been used to model the IETs of relative vorticity extremes \citep{merle, christina}, we consider it here for comparison. 
Although a seasonal time-dependent threshold was used to extract the extreme heat days, some form of seasonality might still be present in the IETs. Therefore, we conduct the Pearson's chi-squared tests separately for each month and grid point, using only the observations from the respective month for the test. In this study, the IET between two successive heat days is assigned to the first of the two days. Under the null hypothesis, a rejection is expected at approximately 5\% of the grid points for each month. The resulting monthly rejection percentages for the eight considered distributions and the two threshold-types are reported in Table~\ref{res_chisq_test_1960_land}. 

\begin{table}[ht]
\centering
\caption{Monthly results of the Pearson's chi-squared goodness-of-fit tests; rejection rates below 5\% are marked in green.}
\begin{tabular}{l l| c c c c c c c c}
  \hline
   & \multicolumn{8}{c}{Rejection Rate (\%)} \\
  Threshold & Month  & ZIDW  & ZINB & ZIGP & ZIG & ZIP & DW & $\delta_0$+WB & FCPP\\
  \hline
  no trend & Jan  & 9.6 & 33.0 & 55.9 & 65.6  & 100.0 & \phantom{0}9.9 & 22.8 & 49.5 \\ 
  
  & Feb  & \textcolor{ForestGreen}{2.5} & 28.1 & 43.0 &  43.0 & 100.0 & \phantom{0}6.2 & 15.4 & 34.0 \\ 
  
  & Mar  & \textcolor{ForestGreen}{2.5} & 36.3 &  46.2 &  44.6 & 100.0 & 12.1 & \phantom{0}8.4 & 26.0\\ 
  
  & Apr  & \textcolor{ForestGreen}{4.0} & 51.1 & 62.3 & 48.7 & 100.0 & 29.8 & 11.7 &  28.2\\
  
  & May  & \textcolor{ForestGreen}{2.4} & 57.3 & 47.8 & 37.2 & 100.0 & 30.9 &  \phantom{0}9.9 & 23.6 \\
  
  & Jun  & \textcolor{ForestGreen}{2.1} & 46.4 & 49.2 &  39.7 & 100.0 & 19.1 &  10.8 & 21.9\\ 
  
  & Jul  & \textcolor{ForestGreen}{1.6} & 44.6 &  46.0 & 49.3 & 100.0 & 15.0 & 12.5 & 31.9  \\
  
  & Aug  & \textcolor{ForestGreen}{2.5}  & 50.5 & 48.8 & 51.5 & 100.0 & 16.5 & 12.1 &  30.6 \\
  
  & Sep  & \textcolor{ForestGreen}{2.1} & 49.2 & 55.8 & 52.6 & 100.0 & 16.1 &  \phantom{0}9.9 & 31.8 \\ 
  
  & Oct  & \textcolor{ForestGreen}{2.4} & 48.8 & 55.5 &  48.7 & 100.0 & 23.2 & \phantom{0}9.2 &  26.5\\
  
  & Nov  & \textcolor{ForestGreen}{1.7} & 26.1 & 56.1 & 43.8 & 100.0 & 10.9 &  \phantom{0}6.6 & 26.0 \\ 
  
  & Dec  & \textcolor{ForestGreen}{4.2} & 37.6 & 52.4 &  63.3 & 100.0 & \phantom{0} 8.4 & 14.6 & 39.2\\ 
  \hline
  trend & Jan  & 6.9 &  29.4 & 48.9 & 54.4 & 100.0 & 10.7 & 21.1 & 43.9\\
  
  & Feb  & \textcolor{ForestGreen}{1.8} &  31.7 &  47.0 & 46.8 & 100.0 & 13.3 & 11.1 &  34.4\\ 
  
  & Mar  & \textcolor{ForestGreen}{2.1} & 36.7 & 39.3 & 41.0 & 100.0 & 17.8 & \phantom{0}9.8 & 28.5 \\ 
  
  & Apr  & \textcolor{ForestGreen}{2.6} & 51.3 & 45.0 &  32.2 &  100.0 & 44.5 & 10.6 & 24.3\\
  
  & May  & \textcolor{ForestGreen}{1.8} & 55.8 & 37.5 & 26.8 & 100.0 & 44.2 & \phantom{0}7.7 & 14.9 \\
  
  & Jun  & \textcolor{ForestGreen}{2.9} & 46.4 & 39.2 &  28.8 & 100.0 & 32.6 & 10.9 &  20.7 \\ 
  
  & Jul  & \textcolor{ForestGreen}{2.5} & 46.4 & 36.7 &  41.0 & 100.0 & 23.4 & 11.7 & 28.8 \\
  
  & Aug  & \textcolor{ForestGreen}{1.6}  & 50.9 & 37.6 & 31.1 & 100.0 & 36.8 & 11.3 & 19.7\\
  
  & Sep  & \textcolor{ForestGreen}{2.6} & 50.4 & 48.3 & 35.0 & 100.0 & 23.1 & 10.2 & 25.2 \\ 
  
  & Oct  & \textcolor{ForestGreen}{3.3} & 40.0 & 42.3 & 37.2 & 100.0 & 27.6 & \phantom{0}9.9 & 26.0 \\
  
  & Nov  & \textcolor{ForestGreen}{0.8} & 22.4 & 43.3 & 36.5 & 100.0 & 17.8 &  \phantom{0}6.9 & 20.8 \\ 
  
  & Dec  & \textcolor{ForestGreen}{2.6} & 35.9 & 41.2 & 46.2 &  100.0 & 18.6 & 11.6 & 30.9 \\ 
  \hline
\end{tabular}
\label{res_chisq_test_1960_land}
\end{table}

For both threshold types, the ZIDW clearly provides the best fit in terms of Pearson's chi-squared tests among all considered distributions. The null hypothesis that the IETs follow a ZIDW distribution is rejected at less than 5\% of the grid points for eleven of the twelve months, the exception being January with 9.6\% (non-trend-adjusted threshold) and 6.9\% (trend-adjusted threshold).
As the IETs are measured in days and therefore lie on a discrete scale, the ZIDW consistently provides a better fit than its continuous counterpart, $\delta_0$+WB, highlighting the advantages of the less commonly used discrete version of the Weibull distribution. The ZIDW also outperforms the DW every month, showing that explicitly addressing zero-inflation improves the fit. For both threshold-types, this is particularly pronounced in April and May, where the rejection rates for the DW are 29.8\% and 30.9\% (non-trend-adjusted threshold) and 44.5\% and 44.2\% (trend-adjusted threshold), respectively. 
The ZIP fails to capture the behavior of the IETs with a rejection rate of 100\% in every month for both threshold-types, reflecting its inability to accommodate heavy-tailed behavior. The ZINB, ZIGP and ZIG also perform poorly, with rejection rates ranging from 26.1\% to 65.6\% (non-trend-adjusted threshold) and 26.8\% to 55.8\% (trend-adjusted threshold).
The FCPP does not adequately capture the behavior of IETs between heat days, with rejection rates of 21.9\%-49.5\% (non-trend-adjusted threshold) and 14.9\%-43.9\% (trend-adjusted threshold). The FCPP, consisting of a Mittag-Leffler distribution mixed with a point mass at zero, describes the heavy upper tail of the IETs well, but fails to capture the high frequency of small, non-zero IETs observed in the data.  

\begin{figure}[!ht]
    \centering
    \includegraphics[width = 0.7\linewidth]{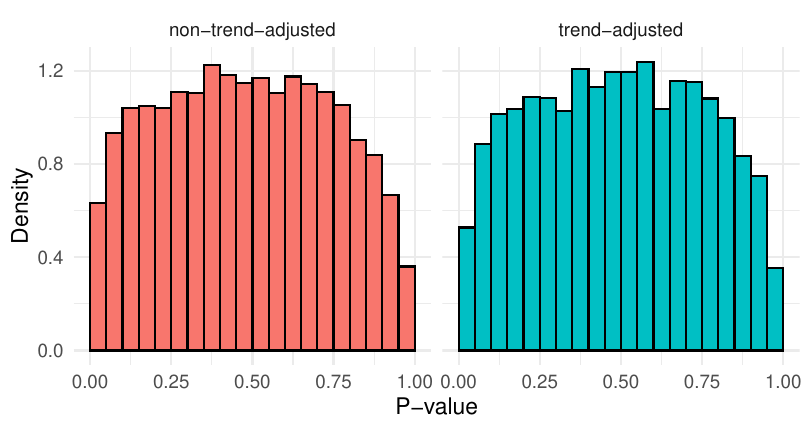}
    \caption{P-values of the Pearson's chi-squared tests for a ZIDW distribution.}
    \label{pvalues_ZIDW}
\end{figure}

In addition to the rejection rates, we study the p-values of the chi-squared tests. Under the null hypothesis of a correctly specified model, the p-values should follow a uniform distribution. Figure~\ref{pvalues_ZIDW} shows the p-values of the Pearson's chi-squared tests for the ZIDW distribution. It can be seen that both small and large p-values occur less frequently than expected, whereas values in the range 0.1 to 0.85 are well represented. Although deviations from the uniform distribution are visible, they occur on both tails of the distribution. Therefore, we do not consider this indicative of substantial model misspecification.

As discussed above, the ZIDW distribution describes the IETs of extreme heat days well. It remains to be assessed whether the duration of extreme heat events, defined as the number of consecutive heat days, is also described adequately under the assumption of independently ZIDW-distributed IETs. 
In this setting, the probability of another heat day occurring on the following day, given that the current day is a heat day, is 
\[\text{P}(\text{IET} = 0) = 1-q(1-\lambda).\] 
Consequently, when assuming independent and ZIDW-distributed IETs, the duration of a heat event follows a geometric distribution with success probability $q (1-\lambda )$. In this relatively simple model, the probability of a heat day occurring on the following day depends only on whether the current day is a heat day, irrespective of the number of preceding consecutive heat days.
Since fewer heat events are observed than single heat days, the rule of thumb for Pearson's chi-squared test, i.e., that no more than 20\% of the bins should have expected frequencies below 5, cannot be satisfied when using monthly data. Hence, we aggregate the data by meteorological season to increase the sample size. However, even with this seasonal aggregation, the rule of thumb is still violated for 4.1-8.4\% (trend-adjusted threshold) and 15.5-22.6\% (non-trend-adjusted threshold) of the grid points.
The results are presented in Table~\ref{res_cl_test_season}. Using thresholds without trend, the null hypothesis is rejected slightly more frequently than expected under the null hypothesis in winter, summer, and fall. However, when trend is included in the threshold, the rejection rates decrease substantially. This is most likely due to an underlying trend in the data, making the earlier observations too different from the later ones. 
For three seasons, the rejection rates fall below the expected 5\% when accounting for a trend, while it is 5.5\% in fall, still indicating an adequate fit. 

\begin{table}[ht]
\centering
\caption{Results of the Pearson's chi-squared goodness-of-fit tests for the heatwave duration; rejection rates below 5\% are marked in green.} 
\begin{tabular}{l| c c c c }
  \hline
   & \multicolumn{4}{c}{Rejection Rate (\%)} \\
  Threshold & Winter (DJF) & Spring (MAM) & Summer (JJA) & Fall (SON) \\
  \hline
  no trend &  6.2 & 6.6 & 6.1 & 5.9 \\
  trend & \textcolor{ForestGreen}{4.9} & \textcolor{ForestGreen}{3.6} & \textcolor{ForestGreen}{4.6} &  5.5 
\end{tabular}
\label{res_cl_test_season}
\end{table}

\begin{figure}[!ht]
    \centering
    \includegraphics[width = 0.7\linewidth]{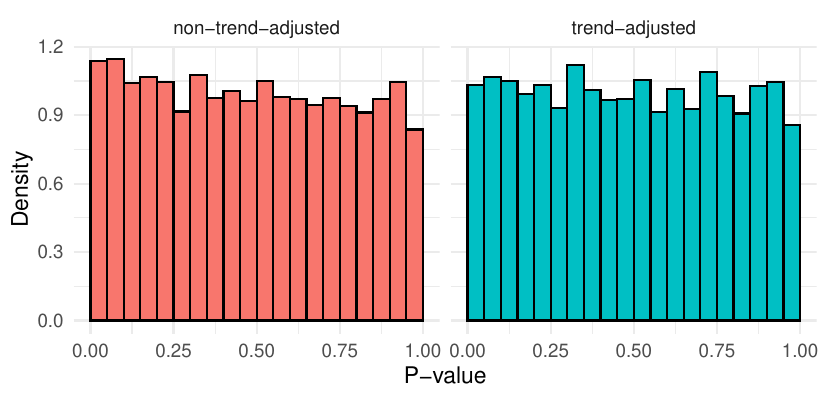}
    \caption{P-values of the Pearson's chi-squared tests for the heatwave duration.}
    \label{pvalues_cl}
\end{figure}

As above, we additionally analyze the p-values of the Pearson's chi-squared tests for the heatwave duration, shown in Figure~\ref{pvalues_cl}. The histograms agree reasonably well with the expected uniform distribution.
Our modeling approach for heat days using a ZIDW distribution assumes independence between the heatwave durations and the IETs between heatwaves.
To assess the validity of this assumption, we compute Spearman's rank correlation between the durations and the IETs between heatwaves. For the trend-adjusted threshold, the Spearman correlations range from $-0.13$ to $0.11$ across all locations, with a median and mean of $-0.02$. For the non-trend-adjusted threshold, a slight tendency towards negative dependence is observed, which can be attributed to the earlier years having longer IETs and shorter heatwaves. However, when considering only the years 2005-2024, the correlations range from $-0.23$ to $0.22$, with a median of $-0.03$ and a mean of $-0.02$. Hence, we do not find evidence of a substantial dependence between heatwave duration and the IETs between heatwaves.
Based on the results presented in this subsection, the ZIDW distribution appears to be appropriate for modeling the IETs of extreme heat days when seasonality is taken into account. 

\subsection{Analysis of temporal changes in the IETs} \label{analysis_trend}

In this section, we compute maximum likelihood estimates of the ZIDW parameters $q$, $\beta$ and $\lambda$ to derive and analyze quantities such as the expected IET and the probability of a zero IET.
To investigate seasonal differences and temporal changes in behavior, the ZIDW distribution is fitted separately for the periods 1960–1992 and 1993–2024, for each of the four seasons. We first examine the IETs derived from the threshold without trend (Section~\ref{analysis_wo_trend}), followed by IETs based on the trend-adjusted threshold (Section~\ref{analysis_with_trend}).

\subsubsection{Threshold without trend} \label{analysis_wo_trend}

\begin{figure}[!ht]
    \centering
    \includegraphics[width = 0.62\linewidth]{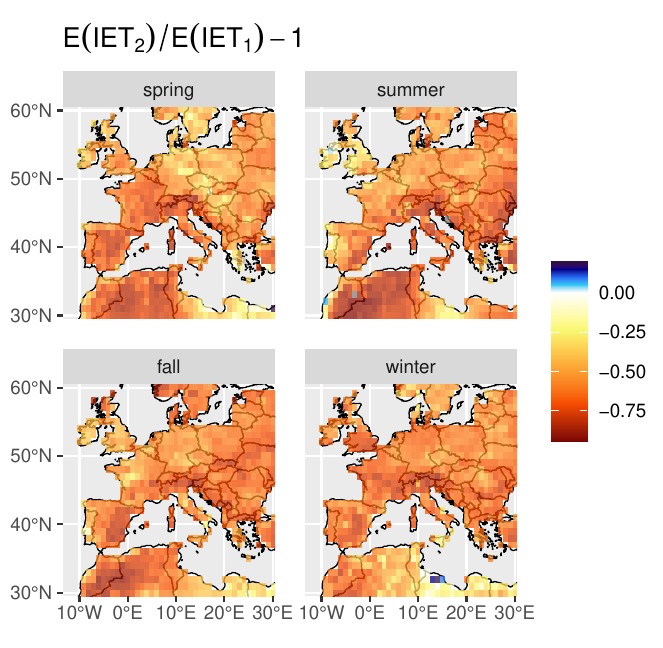}
    \caption{Relative change in estimated expected IET from period 1 (1960-1992) to period 2 (1993-2024), stratified by season; threshold without trend.}
    \label{rel_mean_iet}
\end{figure}

Let $\text{IET}_1$ and $\text{IET}_2$ denote the random IETs between two heat days in the periods 1960-1992 and 1993-2024, respectively. Heat days in both periods are identified using the same locally defined, non-trend-adjusted threshold.
Figure~\ref{rel_mean_iet} depicts the relative change in the estimated expected IET between the two periods, stratified by season, where the expected IETs are derived from the locally fitted ZIDW distributions. For each season, we compute 
\begin{align}\label{EIET_formula}
\frac{\text{E}(\text{IET}_2)-\text{E}(\text{IET}_1)}{\text{E}(\text{IET}_1)} = \frac{\text{E}(\text{IET}_2)}{\text{E}(\text{IET}_1)} - 1 .
\end{align}

Thus, a negative value implies a decrease in the expected IET from the first to the second period, corresponding to more frequent heat days, and is shown in red. All four seasons exhibit similar spatial patterns, and many regions show comparable relative changes in the expected IET across seasons. A pronounced decrease across all seasons is observed in northwestern Africa, Spain, the Alps, southern France and along the Balkan coast. In summer, the western coast of the Black Sea also shows a large relative decrease, whereas the decrease in France stands out in spring and winter. In contrast, smaller relative changes are found in northeastern Africa and Ireland across all seasons, southern Scandinavia (except in fall), as well as in parts of Central Europe in spring.
Overall, the expected IET decreases from the first to the second period at almost all locations, implying a higher frequency of heat days. In all seasons, over 95\% of the coordinate points exhibit a decline in expected IET of at least 25\%, and for more than 70\% of the coordinate points the decline exceeds 50\%. Only a few grid points show an increase in expected IET, specifically one grid point in spring, two in summer, and three in winter.

\begin{figure}[!ht]
    \centering
    \includegraphics[width = 0.62\linewidth]{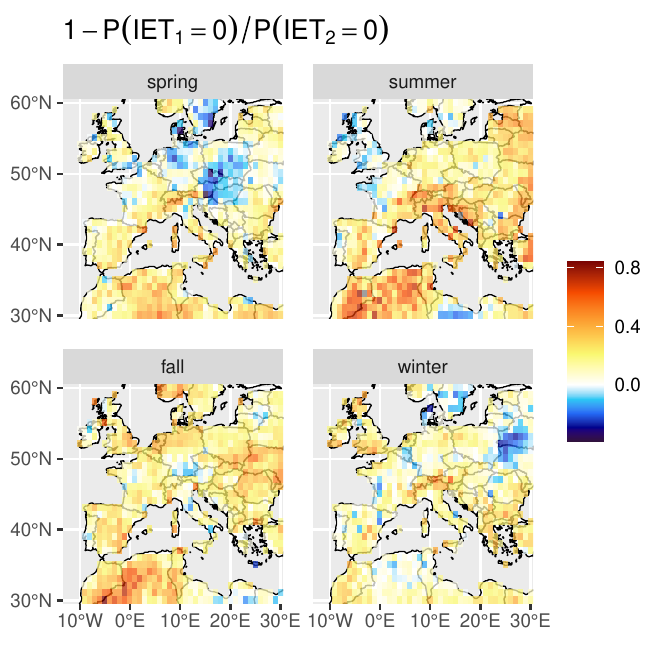}
    \caption{Relative change in the estimated probability of a zero IET from period 1 (1960-1992) to period~2 (1993-2024), stratified by season; threshold without trend.}
    \label{rel_pzero}
\end{figure}

An increase in the number of extreme heat days does not necessarily imply stronger clustering behavior. For instance, heat days could occur more often but relatively regularly spaced instead of clustered. To investigate the clustering behavior of the extreme heat days, we consider the change in the estimated probability of a zero IET between the two periods, stratified by season, again derived from the locally fitted ZIDW distributions. For each season, we calculate 
\begin{align}\label{pzero_formula}
    \frac{\text{P}(\text{IET}_2 = 0)-\text{P}(\text{IET}_1 = 0 )}{\text{P}(\text{IET}_2 = 0)} = 1- \frac{\text{P}(\text{IET}_1 = 0)}{\text{P}(\text{IET}_2 = 0)}.
\end{align}
The second period is chosen as the reference period in the denominator to allow for better visual comparability in the heatmaps. Using the first period as reference would result in a few extremely large values dominating the color scale and obscuring spatial variation across most grid points. The results are shown in Figure~\ref{rel_pzero}. A positive value implies an increase in the probability of consecutive heat days from the first to the second period and is shown in red. This can be interpreted as an indicator of temporal clustering.

While Figure~\ref{rel_mean_iet} shows a general decrease in expected IET across the entire region and all four seasons, the probability of a zero IET and thus the clustering behavior does not admit such a simple and general interpretation. The heatmaps in Figure~\ref{rel_pzero} also exhibit greater seasonal variability, as regions behave differently across seasons.
The largest relative increases in the probability of a zero IET are observed in summer and fall in Morocco and Algeria. In summer, the Alpine region, Italy and Turkey also stand out with pronounced increases. This indicates stronger temporal clustering of heat days and, consequently, longer heatwaves in these regions during the second period. 
A decrease in the probability of a zero IET can be observed as well: in spring predominantly in Sweden and Central Europe (162 grid points total, i.e., 21.4\% of grid point), in winter especially in western Ukraine and Belarus (122 grid points, 16.1\%), in summer in the UK, northwestern France and Libya (65 grid points total, 8.6\%), and in fall in southern Central Europe and northern Libya and Egypt (50 grid points total, 6.6\%). 
Nevertheless, the probability of a zero IET still increases for the majority of locations, indicating an overall growing tendency toward clustering of heat days in Europe and northern Africa. The maximum relative changes observed across the study region are larger in summer and fall than in winter and spring for both the expected IET and the probability of a zero IET.

\subsubsection{Trend-adjusted threshold} \label{analysis_with_trend}

\begin{figure}[!htbp]
    \centering
    \includegraphics[width = 0.62\linewidth]{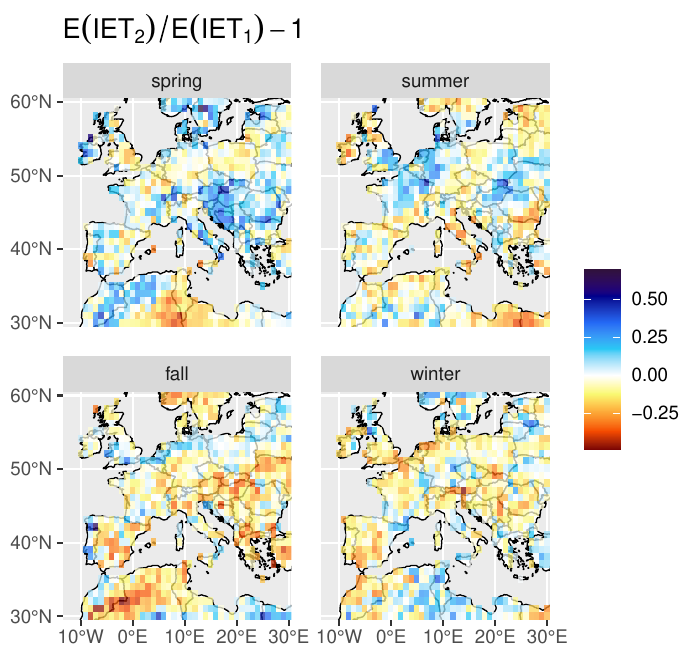}
    \caption{Relative change in expected IET from period 1 (1960-1992) to period 2 (1993-2024) separated by season; trend-adjusted threshold.}
    \label{rel_mean_iet_trend}
\end{figure}

In the following, the analysis of Subsection~\ref{analysis_wo_trend} is repeated, but extreme heat days are now derived from a trend-adjusted threshold. Again, $\text{IET}_1$ and $\text{IET}_2$ denote the random IETs between two heat days in the 1960-1992 and 1993-2024 periods, respectively. 
Figure~\ref{rel_mean_iet_trend} displays the relative change in the estimated expected IET between the two periods, as defined in (\ref{EIET_formula}), stratified by season.
The range of values shows that the maximum relative increase (positive values) is much larger than in the previous section, while the maximum relative decrease (negative values) is smaller in magnitude. 
In addition, the seasonal variability appears to be larger than in Figure~\ref{rel_mean_iet}. For instance, southern Scandinavia shows mostly an increase in expected IET in spring, a decrease in expected IET in summer and fall, and a mixture of both in winter.
Overall, increases in expected IET occur more frequently than in Section~\ref{analysis_wo_trend}, specifically in 419 grid points in spring (55.3\%), 310 in summer (40.9\%), 238 in fall (31.4\%) and 283 in winter (37.3\%). Hence, when adjusting for the general increase in temperature, we do not find a strong tendency toward a decreasing expected IET. This result is not unexpected, as the trend-adjusted threshold ensures that approximately the same number of extreme heat days occurs in both periods.

\begin{figure}[!htbp]
    \centering
    \includegraphics[width = 0.62\linewidth]{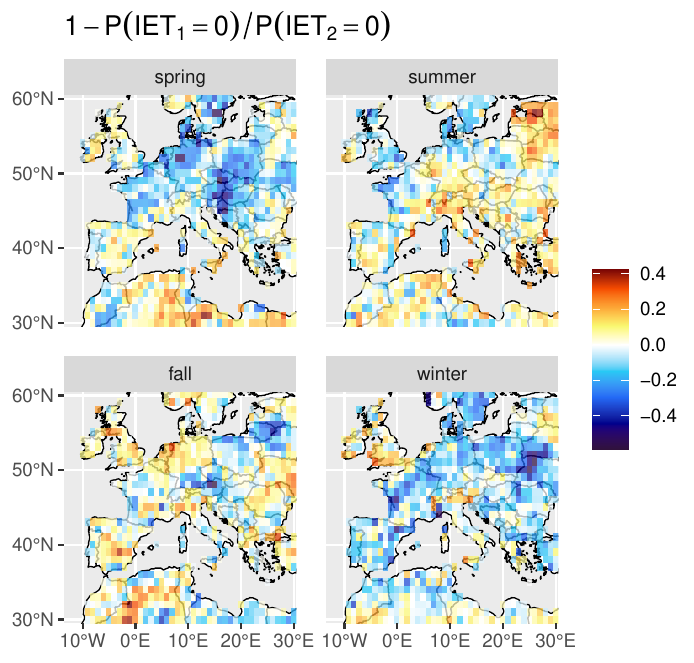}
    \caption{Relative change in the estimated probability of a zero IET from period 1 (1960-1992) to period 2 (1993-2024) separated by season; trend-adjusted threshold.}
    \label{p_zero_trend}
\end{figure}

Figure~\ref{p_zero_trend} depicts the relative change in the probability of a zero IET, as defined in (\ref{pzero_formula}), stratified by season.
An increase in the probability of zero can be seen at 322 locations in spring (42.5\%), 448 in summer (59.1\%), 439 in fall (57.9\%) and only at 226 locations in winter (29.8\%). This suggests that the tendency towards temporal clustering of extreme heat days does not generally increase when adjusting for the overall increase in temperature. Especially in winter, the clustering behavior of heat days predominantly decreased, with notable exceptions in the British Isles and northern Italy.
In summer and fall, increased clustering of heat days, associated with longer heatwaves, is observed in southern Europe and northern Africa, particularly across the Iberian Peninsula and the Mediterranean region, as well as in parts of eastern Europe. Additional increases are observed in northeastern Europe during summer and in Central Europe during fall.

Overall, the results indicate an increase in temperature extremes throughout the study region, see Figure~\ref{threshold_epanechnikov}, accompanied by a stronger temporal clustering of heat days for most locations (Figure~\ref{rel_pzero}). However, when the definition of extreme heat is adjusted to account for temporal changes, no uniform change in the clustering behavior of extreme heat days is observed in the region. In this setting, Figure~\ref{p_zero_trend} shows that in winter the estimated probability of a zero IET is predominantly decreasing, whereas the other seasons show a relatively balanced ratio between locations with an increasing and decreasing estimated probability of a zero IET.

\subsection{European summer 2026} \label{2026}

\begin{figure}[!htbp]
    \centering
    \includegraphics[width = 0.62\linewidth]{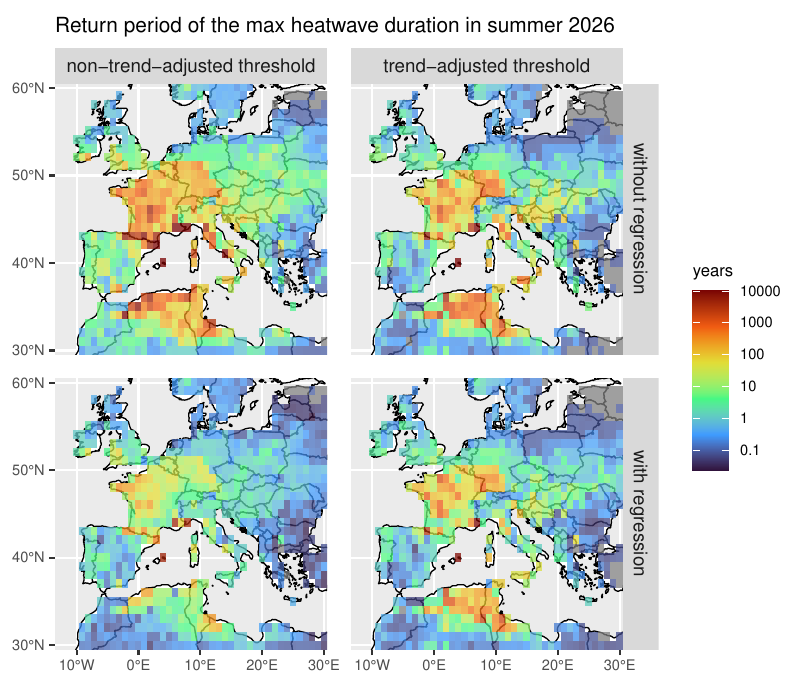}
    \caption{Estimated return periods of the maximum heatwave duration in summer 2026 in years, stratified by threshold type and whether a regression model was used; values exceeding 10{,}000 are truncated at 10{,}000 and locations without heat events are colored gray.}
    \label{return_period26}
\end{figure}

Summer 2026 (June to August) was the warmest summer on record for western Europe and the third-warmest summer on record for Europe overall \citep{august_2026}. Several national temperature records were broken, for instance in Germany and Denmark \citep{june_2026}.
In this subsection, we assess whether summer 2026 was also unusual in terms of its heatwave durations. To this end, we consider the daily temperature maxima from June 1 to August 31, 2026, of the ERA5 reanalysis data.
Heat days are identified using both threshold types, and the longest heatwave duration observed in summer 2026 is extracted for each location. The 2026 trend-adjusted threshold is approximated using the corresponding threshold for 2024.
For each maximum duration $d_\text{max}$, we compute the corresponding return period under the assumption of a ZIDW distribution for the IETs between heat days, which is fitted to the summer data, separately for the trend-adjusted and non-trend-adjusted thresholds. As explained in Section~\ref{goodness-of-fit}, assuming a ZIDW distribution for the IETs between heat days yields a geometric distribution for the heatwave duration $D$. The return period $T_D(d_\text{max})$ is defined as the expected time until a heatwave of at least the duration $d_\text{max}$ occurs and is given by 
\[T_D(d_\text{max}) = \frac{1}{\text{P}(D\geq d_\text{max} ) }\cdot [ \text{E}(D) +\text{E}(\text{IET}\;| \;\text{IET} > 0) ],\]
converted to years.
Additionally, we fit regression models with temporal trends in the parameters for both threshold types and  compute the return periods of the maximum heat wave durations. For each location and threshold type, we consider four specifications for each of the three parameters $q$, $\beta$ and $\lambda$: constant, linear, quadratic, and cubic, which yields $4^3 = 64$ possible model specifications. For the return period analysis, we select the model yielding the shortest return period for summer 2026 at each location and for each threshold type, as this provides a conservative assessment of the extremeness of the observed heatwave duration.
The results, stratified by threshold type and whether a regression model was used, are displayed in Figure~\ref{return_period26}. Six values exceeding 10{,}000 years are truncated at 10{,}000 to improve visual comparability and locations without heat events in summer 2026 are shown in gray.

It is evident that summer 2026 was considerably more unusual in western and central Europe and parts of northern Africa than in eastern Europe, regardless of threshold type or whether regression was used. Especially France, southern Germany, northern Algeria and Tunisia exhibit high estimated return periods. Without regression, the return periods exceed 1{,}000 years at 29 and 16 locations for the non-trend-adjusted and trend-adjusted thresholds, respectively. 
This is the case despite partly using a trend-adjusted threshold in the heat day definition, highlighting the severity of this summer in western Europe. 
However, incorporating regression models reduces the estimated return periods, with only three and nine locations exceeding 1{,}000 years for the non-trend-adjusted and trend-adjusted thresholds, respectively.
The reduction is particularly pronounced for the non-trend-adjusted threshold, for which regression yields the lowest estimated return period among the four approaches at 450 of the 748 locations with observed heat events during summer 2026.
Among the regression models yielding the lowest estimated return periods for the non-trend-adjusted threshold, a temporal trend in $q$ is included in 727 cases, of which 595 are statistically significant at the 5\% level. Of these significant trends, 543 are decreasing, implying an increase in the probability of small IETs when $\beta$ and $\lambda$ remain constant. Temporal trends in $\beta$ and $\lambda$ are included in 569 and 719 cases, respectively, with statistically significant trends observed in only 151 and 112 cases, respectively. Of these significant trends, 84 in $\beta$ are negative, indicating a reduction in the probability of large IETs, whereas 102 in $\lambda$ are positive, indicating an increase in zero-inflation. A similar pattern is observed for the trend-adjusted threshold. A temporal trend in $q$ is included in 713 cases, of which 417 are statistically significant; 333 of these significant trends are negative. In contrast, temporal trends in $\beta$ and $\lambda$ are included in 692 and 706 cases, respectively, but are statistically significant in only 115 and 100 cases. Of these significant trends, 88 in $\beta$ are negative, whereas 83 in $\lambda$ are positive.

It should be noted that the interpretation of heatwave durations from the ZIDW model relies on the assumption of independently distributed IETs. While the goodness-of-fit results in Section~\ref{goodness-of-fit} provide support for this approximation, temporal dependence may affect the probabilities of long heatwaves and the corresponding return periods.
    \section{Discussion} \label{sec:conclusion}

In this study, we analyzed inter-exceedance times (IETs) between heat days in mid-latitude Europe and northern Africa, where heat days are defined by exceedances of local calendar-day-based 95\% quantiles. 
By modeling IETs directly, we provide a unified framework that captures both the duration of heatwaves and the waiting times between them within a single model. Across all considered candidate distributions, the zero-inflated discrete Weibull (ZIDW) distribution consistently provided the best fit, measured by the Pearson's chi-squared goodness-of-fit tests. The ZIDW distribution provides a suitable model for heat days, as it simultaneously captures excess zeros, interpreted here as consecutive heat days clustering into a heatwave, and over-dispersion.

Heat days were identified using both trend-adjusted and non–trend-adjusted seasonal thresholds, since the seasonal variation in the thresholds allows the analysis of unusually warm days outside summer. The non–trend-adjusted threshold is suitable for evaluating general changes in the occurrence of extreme heat days, whereas trend-adjusted thresholds enable the analysis of changes after accounting for the overall increase in temperature.
The ZIDW distribution was fitted separately for each location in the study region, stratified by season, for the periods 1960–1992 and 1993–2024.
When using non–trend-adjusted thresholds, the expected IET decreases substantially across nearly the entire study region, indicating a strong increase in the frequency of extreme heat days. However, the clustering behavior, as measured by the probability of zero IETs, shows a more nuanced spatial and seasonal pattern. While an overall increase in clustering is observed in many regions, a few regions exhibit decreasing clustering, e.g., central and northern Europe in spring.
The observed patterns change markedly when trend-adjusted thresholds are used. After accounting for the general increase in temperature in the threshold, the strong and spatially coherent decrease in expected IET largely disappears. Instead, both increases and decreases in expected IET are observed, depending on the region and season. Similarly, the previously observed widespread increase in clustering is no longer present. In particular, winter shows a tendency toward decreasing clustering, while the other seasons exhibit a more balanced pattern between increasing and decreasing clustering.

Furthermore, the severity of summer 2026 was assessed in terms of heatwave durations.
To this end, the longest heatwave observed during summer 2026 was identified for each location and threshold type. 
In addition to fitting a ZIDW model with constant parameters, we fitted regression models incorporating temporal trends in the parameters for both threshold types and the corresponding return periods were estimated for both model approaches and threshold types.
Based on these modeling approaches, the return periods of the longest heatwaves observed in summer 2026 exceeded 100 years at many locations, particularly in France, Tunisia, northern Algeria and southern Germany. High return periods were observed even when using a trend-adjusted threshold for the heatwave definition or incorporating temporal trends in the model parameters through regression, highlighting the severity of summer 2026 in western Europe and parts of northern Africa. In contrast, relatively short return period estimates in eastern Europe suggest that summer 2026 was not particularly unusual in this region.

    \section*{Acknowledgements}
    This research has been funded by the German Federal Ministry of Research, Technology and Space (BMFTR) within the subproject B3.3 (project number 01LP2323K) of the integrated project “Climate Change and Extreme Events – ClimXtreme Module B Statistics Phase II”. The authors gratefully acknowledge the computing time provided on the Linux HPC cluster at TU Dortmund University (LiDO3), partially funded in the course of the Large-Scale Equipment Initiative by the German Research Foundation (DFG) as project 271512359.
    The authors also express their gratitude to Joaquim Pinto, Karlsruhe Institute of Technology (KIT), for their valuable suggestions and insightful discussions. 
    During the preparation of this manuscript, we used the ChatGPT-4o model developed by OpenAI for minor language editing to improve readability. After using this tool, the authors carefully reviewed and edited the content as needed and take full responsibility for the content of the manuscript.

    \section*{Disclosure statement}
    The authors report that there are no competing interests to declare.

    \section*{Data Availability Statement}
    The ERA5 reanalysis data are openly available through the Copernicus Climate Data Store at \url{https://cds.climate.copernicus.eu/datasets}.

	\printbibliography[heading=bibintoc, title={References}]
    \appendix
    \numberwithin{equation}{section}
    \renewcommand{\theequation}{A-\arabic{equation}}

\end{document}